\documentclass[aps,twocolumn,showpacs,preprintnumbers,nofootinbib,superscriptaddress]{revtex4-1}

\usepackage{amsmath,amssymb,bm}
\usepackage[dvips]{graphicx}
\usepackage{hyperref}
\usepackage{yfonts}
\usepackage{color}
\newcommand{\beq}{\begin{eqnarray}}
\newcommand{\eeq}{\end{eqnarray}}

\renewcommand\d{\partial}

\begin{document}

\title{Magnetars and the Chiral Plasma Instabilities}

\author{Akira Ohnishi}
\affiliation{Yukawa Institute for Theoretical Physics, Kyoto University, 
Kyoto 606-8502, Japan}

\author{Naoki Yamamoto}
\affiliation{Maryland Center for Fundamental Physics, 
Department of Physics, University of Maryland,
College Park, Maryland 20742-4111, USA}

\begin{abstract}
We propose a possible new mechanism for a strong and stable magnetic field of 
compact stars due to an instability in the presence of a chirality imbalance of 
electrons---the chiral plasma instability. A large chirality imbalance of electrons 
inevitably occurs associated with the parity-violating weak process during core 
collapse of supernovae. We estimate the maximal magnetic field due to this 
instability to be of order $10^{18}$ G at the core. This mechanism naturally generates 
a large magnetic helicity from the chiral asymmetry, which ensures the stability of the
large magnetic field.
\end{abstract}
\pacs{97.60.Jd, 12.15.Ji, 52.35.-g}
\maketitle

\begin{figure}[t]
\begin{center}
\includegraphics[width=8cm]{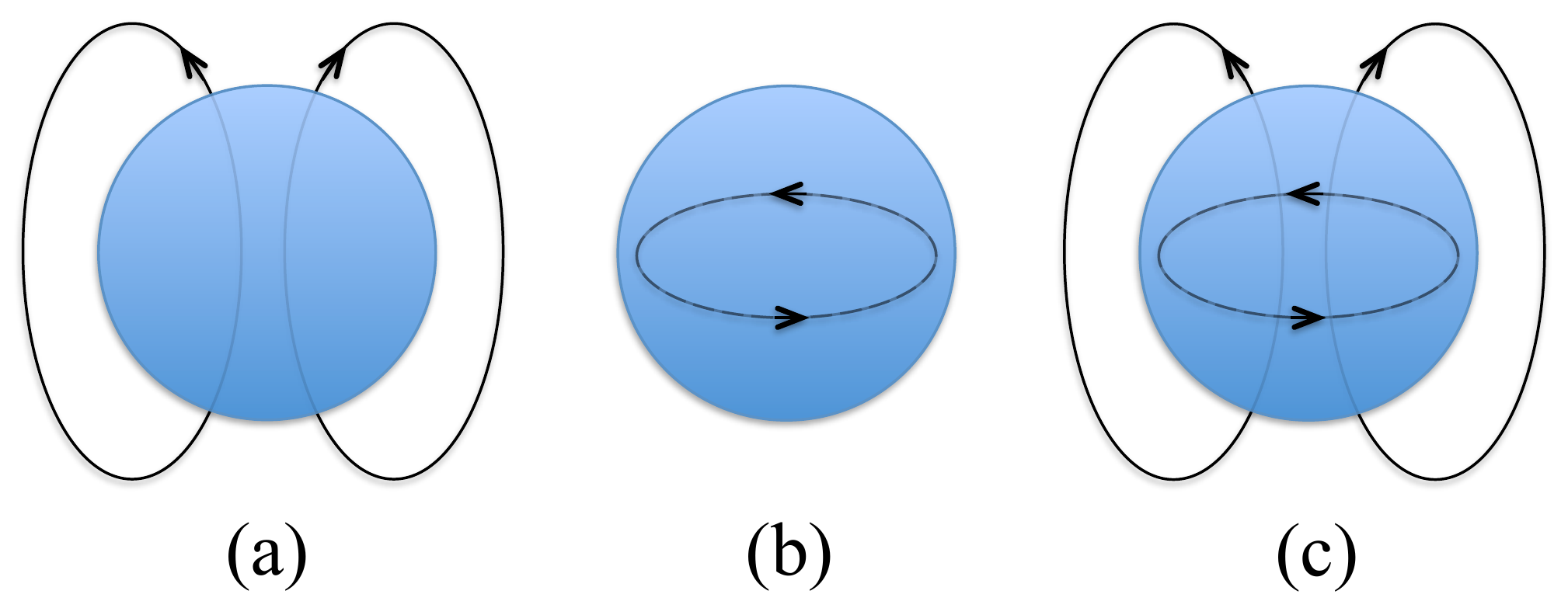}
\end{center}
\vspace{-0.5cm}
\caption{Configurations of magnetic fields in magnetars: (a) poloidal, (b) toroidal, 
and (c) linked poloidal-toroidal magnetic fields.}
\label{fig:star}
\end{figure}

\emph{Introduction.}---%
The origin of compact stars with the most powerful magnetic field ($\sim 10^{15}$ G 
on the surface) in the Universe, called magnetars \cite{Magnetar}, 
is a mystery in astrophysics.
Examples of the possible mechanisms include the fossil field or dynamo hypothesis 
among others \cite{Harding:2006qn, Spruit:2007bt}. However, these mechanisms have an important 
problem that a strong magnetic field produced cannot be sustained for a long time scale
\cite{Harding:2006qn, Spruit:2007bt}. Indeed, a purely poloidal magnetic field 
[depicted in Fig.~\ref{fig:star}(a)] typically considered is known to be unstable \cite{Wright}, 
as numerically confirmed by magnetohydrodynamics \cite{Braithwaite:2005md}. 
For other issues in these mechanisms, see, e.g., Ref.~\cite{Spruit:2007bt}.

It is suggested that \emph{if} nonzero magnetic helicity 
\beq
\label{H}
{\cal H} = \int d{\bm x}\, {\bm A} \cdot {\bm B}
\eeq
(where ${\bm B}$ and ${\bm A}$ are the magnetic field and vector potential) is 
produced at the initial configuration for some reason, it can make the 
magnetic field stable \cite{Spruit:2007bt}. 
This is because ${\cal H}$ is proportional to the Gauss linking number of the 
magnetic flux tubes
and serves as an approximate conserved quantity \cite{H}.%
\footnote{Magnetic helicity is a conserved quantity for a perfect conductor
and is conserved approximately with finite conductivity; see the discussion below.} 
For example, linking of poloidal and toroidal magnetic fields [depicted in Fig.~\ref{fig:star}(c)]
has a nonzero magnetic helicity and can exist stably \cite{Prendergast}. However, the 
origin of the magnetic helicity itself remains to be understood as well (see also below).

In this paper, we propose a new mechanism for a strong \emph{and} stable magnetic
field in magnetars due to a novel instability in the presence of an imbalance 
between right- and left-handed electrons---the chiral plasma instability.
The chiral plasma instability was recently found in the context of electromagnetic and 
quark-gluon plasmas  \cite{Akamatsu:2013pjd} based on chiral kinetic theory \cite{Son:2012wh}.  
A related instability had been previously argued for the electroweak theory
\cite{Redlich:1984md, Rubakov:1985nk}
and for the primordial magnetic field in the early Universe \cite{Joyce:1997uy}.
This instability appears somewhat similar to the Rayleigh-Taylor instability that occurs
in the presence of a density imbalance of two fluids at an interface. However, the former 
is remarkable in that it is a consequence of relativistic and quantum effects related 
to quantum anomalies \cite{Adler} unlike the latter.

Our mechanism for a strong and stable magnetic field is based only on the chirality 
asymmetry of electrons that is inevitably produced in the parity-violating weak 
process (electron capture) during core collapse of supernovae. 
The energy of the chirality imbalance is converted to a large magnetic field by the 
chiral plasma instability. Furthermore, we show that it naturally generates a large 
magnetic helicity at the same time. 
To the best of our knowledge, this is the only microscopic mechanism to create a 
magnetic helicity in magnetars. Note that our mechanism does not require any exotic 
hadron or quark phases inside compact stars under discussion, such as ferromagnetic 
nuclear matter \cite{nuclear}, ferromagnetic quark matter \cite{Tatsumi:1999ab}, pion 
domain walls \cite{Son:2007ny}, and so on, which are essential in some of previous 
suggestions for the origin of magnetars.

In the following, we neglect the electron mass $m_e$, as electrons may be 
regarded as ultrarelativistic at high density, $\mu_e \gg m_e$.
We will discuss the possible effects of $m_e$ later on.

\emph{Chiral Plasma Instabilities.}---%
Let us briefly review the physical argument of the chiral plasma instability 
(see Ref.~\cite{Akamatsu:2014yza} for the detail). For simplicity, we here ignore 
the effect of dissipation 
which is not essential to understand the instability itself.

\begin{figure}[t]
\begin{center}
\includegraphics[width=4cm]{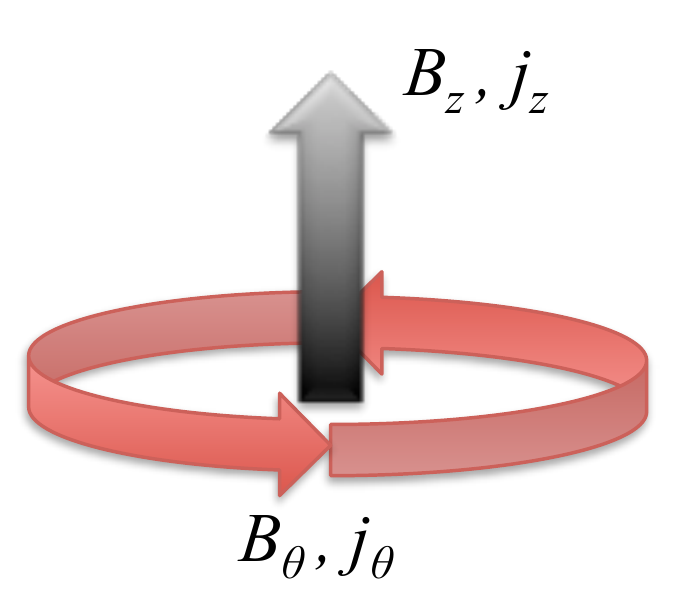}
\end{center}
\vspace{-0.5cm}
\caption{Physical picture of the chiral plasma instability.}
\label{fig:CPI}
\end{figure}

Suppose there is a homogeneous chiral asymmetry between right- and 
left-handed electrons in the core of compact stars, which we parametrize 
by a chiral chemical potential $\mu_5 \equiv (\mu_R - \mu_L)/2$. 
(We shall give an estimate of $\mu_5$ just after the onset of core collapse of supernovae later.)
Let us consider a perturbation of a small magnetic field $B_z$ with wavelength $\lambda$ 
in a cylindrical coordinate $(r, \theta, z)$ 
(see Fig.~\ref{fig:CPI}). In the presence of $\mu_5$, this magnetic field leads to an 
electric current in the $z$ direction (called the chiral magnetic effect) \cite{Vilenkin:1980fu}:
\beq
\label{CME}
j_z = \frac{2 \alpha}{\pi} \mu_5 B_z,
\eeq
where $\alpha$ is the fine structure constant. 
Intuitively, this current can be understood as follows: to minimize the energy of 
the system, the spin of an electron is aligned in the same direction as $B_z$.
Remembering the definition of chirality, this means that momentum of 
a right- (left-)handed fermions is in the same (opposite) direction as $B_z$,
and so is the electric current. Hence, a net electric current flows in the direction 
of $B_z$.

Now Amp\`ere's law states that the current (\ref{CME}) leads to a magnetic field in 
the $\theta$ direction at a distance $R \sim \lambda$ as $B_{\theta} = \pi \lambda^2 j_z/(2\pi R)$.
This in turn induces the current due to the chiral magnetic effect:
\beq
j_{\theta}(R) = \frac{2 \alpha}{\pi} \mu_5 B_{\theta} 
= \left(\frac{2 \alpha \mu_5 \lambda}{\pi} \right)^2 \frac{1}{2 R} B_z.
\eeq
According to Amp\`ere's law again, this current gives rise to a magnetic field 
in the $z$ direction as
\beq
B'_z = \int dR\, j_{\theta}(R) \sim \left(\frac{2 \alpha \mu_5 \lambda}{\pi} \right)^2 B_z.
\eeq
So if $\lambda \gtrsim (\alpha \mu_5)^{-1}$, it follows that $B'_z > B_z$:
the original magnetic field gives a positive feedback to itself, and 
it grows exponentially. This is the chiral plasma instability.
This unstable mode then reduces $\mu_5$ so that the instability is attenuated 
\cite{Akamatsu:2013pjd, Akamatsu:2014yza}.

One also finds that this instability generates not only a poloidal magnetic field in the 
$z$ direction but also a toroidal magnetic field in the $\theta$ direction; the resulting 
configuration has a finite magnetic helicity. 
Later, we will estimate the magnitude of magnetic helicity from the helicity conservation.

\emph{Estimate of the chiral imbalance of electrons.}---%
How large can a magnetic field be due to this mechanism inside compact stars?
From now on, we shall provide its estimate based on the neutron density
generated during core collapse of supernovae. The core of a compact star is
almost ``neutronized'' at this stage via the parity-violating weak process, and 
as a result, the largest chiral asymmetry of electrons is created during its evolution \cite{mu5}.

We note that our estimate for the magnetic field below assumes a number of 
simplifications, so it should be regarded as schematic.
Nonetheless, it turns out that the maximal magnetic field due to the chiral
plasma instability can be of order $10^{18}$ G at the core (with a possible small 
deviation due to the uncertainty of the prefactor), and it might be sufficient enough 
to explain the large magnetic field $\sim 10^{15}$ G on the surface. 
Our estimate can, in principle, be made more realistic by including various
complications that we will discuss later.

The chirality imbalance of electrons is produced via electron capture inside a core,
\beq
\label{capture}
p + e_L^- \rightarrow n + \nu_L^e,
\eeq
where the subscript $L$ stands for left-handedness. 
Here only left-handed fermions are involved, as it is described by the V$-$A type 
weak interaction.
Its inverse process also exists, and reduces the number of neutrons and the chirality 
imbalance. Note that the other possible processes 
(thermal neutrino emission) \cite{Shapiro}
do not change the neutron number nor the chirality imbalance on average. 

That huge neutrons are produced in these processes (to form a \emph{neutron} 
star eventually) means that huge left-handed electrons are ``eaten" by  protons,
leading to the Fermi surface imbalance, $\mu_R > \mu_L$, or a nonzero chiral 
chemical potential for electrons, $\mu_5 \equiv (\mu_R - \mu_L)/2 > 0$.
Because the number of neutrons produced in this process is equal to the number 
difference between right- and left-handed electrons, $N_5$, we have
\beq
\label{n5}
n_5 \approx \Delta n_{\rm n},
\eeq
where $n_5$ is the chiral density of electrons and $\Delta n_{\rm n}$
is the increased neutron density by electron capture. Considering the neutron density
inside a neutron star, it is reasonable to take 
$\Delta n_{\rm n} \sim (0.1$--$1) \ {\rm fm}^{-3}$ at the core.
In the natural units $\hbar = c = 1$, $\Delta n_{\rm n} \sim 0.1$--$1\Lambda^3$, where 
we introduced the mass scale $\Lambda = 200 \ {\rm MeV}$ for later convenience.
In the following, we use the natural units unless otherwise stated. 

The chiral number density is expressed by the chemical potentials as
\beq
\label{n5-mu5}
n_5 = \frac{\mu_5}{3\pi^2}(\mu_5^2 + 3 \mu^2)
\eeq
at sufficiently low temperature, where 
$\mu \equiv (\mu_R + \mu_L)/2$ is the chemical potential associated with 
U(1) (vector-like) particle number. We here ignored the contribution of the temperature
$T$ to the density, because $T$ is at most $\sim 10^{10}$ K ($\sim 10^{-2} \Lambda$) 
at the core, and is negligibly small compared with the contributions of $\mu$ and $\mu_5$
(as will be justified below). Recalling that the typical electron chemical potential at the 
core is $\mu \lesssim \Lambda$, one finds 
\beq
\mu_5 \sim \Lambda.
\eeq

It should be remarked that $\mu_5$ we obtained is the total (or the time-integrated)
chiral chemical potential during core collapse. In reality, the production of $\mu_5$ by
the process (\ref{n5-mu5}) occurs simultaneously with the reduction of $\mu_5$ by the 
chiral plasma instability 
(and with the reduction by the electron mass $m_e$, which we shall argue later).

\emph{Estimate of magnetic fields and magnetic helicity.}---%
As explained above, the state with nonzero $\mu_5$ is unstable and decays 
rapidly by converting it to a magnetic field due to the chiral plasma instability. 
Assuming this state will decay into a state with $\mu_5 \sim 0$ at saturation, 
one can estimate the magnitudes of a resulting magnetic field and magnetic 
helicity from the energy and helicity conservations \cite{Akamatsu:2013pjd}.

The energy conservation requires that the energy density of electrons due to 
the chiral asymmetry, 
\beq
\Delta E = \frac{1}{4\pi^2} (\mu_5^4 + 6 \mu_5^2 \mu^2)
\eeq 
is equal to that of the magnetic field $(1/2)\Delta B_{\rm inst}^2$. 
One can thus estimate the maximal magnetic field generated by this instability as
\beq
\label{B}
B_{\rm max} \sim \Lambda^2 \sim10^{18} \ {\rm G},
\eeq
assuming no dissipation of energy and perfect conversion efficiency
(see the discussion below). We notice here that the magnitude of the magnetic field 
is the QCD scale through the relation (\ref{n5}), although the magnetic field itself is 
produced by the electroweak dynamics. 
We can translate $B_{\rm max}$ at the core into the magnetic field on the surface 
from the conservation of a magnetic flux:
\beq
B_{\rm surface} \approx  \left(\frac{R_{\rm core}}{R_{\rm star}} \right)^2 B_{\rm max},
\eeq
where  $R_{\rm core}$ and $R_{\rm star}$ are the radii of the core and the neutron 
star itself, respectively; e.g., when $R_{\rm core}/R_{\rm star} \sim 10^{-1}$, the maximal 
surface magnetic field is $B_{\rm surface} \sim 10^{16} \ {\rm G}$.

On the other hand, the helicity conservation reads
\beq
\label{helicity_cons}
\frac{d}{dt}\left(N_5 + \frac{\alpha}{\pi}{\cal H} \right)=0,
\quad  N_5 = \int d{\bm x}\, n_5,
\eeq
where $N_5$ is the global chiral charge of electrons and ${\cal H}$
is the magnetic helicity (also called Chern-Simons 
number in particle physics and mathematics) defined in Eq.~(\ref{H}). 
In passing, we note that Eq.~(\ref{helicity_cons}) is the global version of 
the anomaly relation in quantum electrodynamics \cite{Adler}:
\beq
\d_{\mu} j^{\mu 5} = \frac{2 \alpha}{\pi}{\bm E} \cdot {\bm B},
\eeq
where $j^{\mu 5} = \bar e \gamma^{\mu} \gamma^5 e$ is the 
axial current for electrons. 
From the helicity conservation, one obtains the magnetic helicity at saturation as
\beq
\Delta {\cal H} = - \frac{\pi}{\alpha}\Delta N_5 \sim -\frac{1}{\alpha} V \Lambda^3
\eeq
with $V \approx 4\pi R_{\rm core}^3/3$; so a large magnetic helicity is naturally produced 
as a consequence of the chiral plasma instability, which then ensures the stability of
the strong magnetic field. The detailed configuration of the magnetic field with a 
large magnetic helicity is under study \cite{Akamatsu2}.

We note that any microscopic process concerning the electromagnetic and strong 
interactions respects parity and cannot generate a parity-odd magnetic helicity; 
microscopically, a parity-odd quantity can originate from the parity-violating weak interaction alone. 
However, the weak interaction violates parity in the fermionic sector (leptons), so it 
cannot generate magnetic helicity directly. It is this chiral plasma instability that converts 
the parity-odd chiral asymmetry in the fermionic sector to the parity-odd magnetic helicity 
in the gauge sector. 

Note also that the interior of a star could acquire the magnetic helicity \emph{macroscopically} 
by accident, by losing helicity through the surface in its evolution. 
However, no such a evidence was observed in magnetohydrodynamics
for a specific initial configuration with ${\cal H}=0$ \cite{Braithwaite:2005md}.
At least, our mechanism seems the only to generate magnetic helicity microscopically.

\emph{Discussion.}---%
Let us discuss several possible effects we have ignored above, which can modify
our simple estimate (\ref{B}). One potentially important effect is the electron mass 
$m_e$ which also reduces the chiral asymmetry; for a given $\mu_5$ generated by the 
process (\ref{capture}), the chirality flipping rate due to $m_e$ is 
$\Gamma_{\rm mass} \sim \alpha^2 (m_e/\mu)^2 \mu_5$ \cite{Joyce:1997uy}. 
The time scale $\sim \Gamma_{\rm mass}^{-1}$ is then much larger than that of the 
chiral plasma instability, $\tau_{\rm inst} \sim (\alpha^2 \mu_5)^{-1}$ \cite{Akamatsu:2013pjd}, 
by a factor of $(\mu/m_e)^2 \gg 1$; hence, the effects of $m_e$ is expected to be minor.
Another (possibly more) important effect is the conversion efficiency of the chirality imbalance 
into the magnetic energy which could be less than 100\%. For example, the conductivity 
$\sigma$ dissipates the energy and makes the magnetic field in Eq.~(\ref{B}) smaller. 
Also, the nuclear (or QCD) dynamics inside a star may interfere with the electron 
dynamics and could reduce the magnetic field.
Although evaluation of these effects is not easy, it is not entirely unreasonable 
to expect that the magnetic field induced by the chiral plasma instability can occupy a 
nonnegligible fraction of the gigantic magnetic field of magnetars.

Finally, we comment on the possible evolution of the large magnetic field after 
the birth of magnetars by our mechanism.
Remember that the magnetic helicity ${\cal H}$ is a strict conserved quantity without 
dissipation. In reality, the medium has a conductivity $\sigma$ so that ${\cal H}$ 
is conserved approximately; it is the finiteness of $\sigma$ that allows magnetic flux 
tubes to reconnect, which results in the decrease of ${\cal H}$. 
Therefore, one expects that magnetic fields decay slowly by dissipation and the 
reconnection which could manifest themselves in the form of giant outbursts.

\emph{Conclusion.}---%
In conclusion, we proposed a possible new mechanism for a strong magnetic field
with magnetic helicity in magnetars due to the chiral plasma instability. 
Our mechanism is based only on the chirality imbalance of electrons that is
necessarily produced during core collapse. The maximal magnetic field is estimated 
as $\sim 10^{18}$ G at the core which may be sufficient to explain the magnetic 
field $\sim 10^{15}$ G on the surface. This large magnetic field is a macroscopic 
consequence of the relativistic and quantum effects. More realistic calculations, e.g., 
using magnetohydrodynamics for chiral plasmas, are necessary to reach a definite 
conclusion. We defer these calculations to future work.

We thank Y.~Akamatsu, P.~Bedaque, T.~Enoto, S.~Mahmoodifar, K.~Sumiyoshi, 
and Y.~Suwa for useful discussions. 
N.~Y. was supported by JSPS Research Fellowships for Young Scientists when this 
collaboration started.

\end{document}